\documentclass{ws-procs9x6}

\begin{document}        

\title{Who Confines Quarks?  \\ 
- On Non-Abelian Monopoles and  \\
 Dynamics of Confinement     }

\author{K. Konishi} 
  
\address{Dipartimento di Fisica,``E. Fermi"\\
Universit\`a di Pisa, \\ 
Via Buonarroti, 2, Ed. C\\
56127 Pisa, Italy \\ 
E-mail: konishi@df.unipi.it}


\maketitle

\abstracts{
The role non-Abelian magnetic monopoles play    in the dynamics of confinement is discussed  by examining carefully 
 a class of  
 supersymmetric gauge theories 
as  theoretical laboratories.  In particular, in the so-called 
$r$-vacua of  softly broken ${ N}=2$ supersymmmetric  $SU(n_c)$  QCD,  the Goddard-Olive-Nuyts-Weinberg  monopoles
appear as the dominant  low-energy effective degrees of freedom.  Even more interesting is the    physics of  confining vacua which
are deformations  of nontrivial superconformal theories.     We argue that in  such cases, occurring    in the  $r= {n_f \over 2}$ vacua of  
$SU(n_c)$ theories  or in all of confining vacua of $USp(2n_c)$ or $SO(n_f)$ theories with massless flavors,   a new mechanism of confinement 
involving strongly interacting non-Abelian magnetic monopoles is at work.    }

\section{Confinement as  a dual superconductor? }

The basic issue underlying  the problem of confinement and dynamical symmetry breaking  in QCD  
is the nature of the effective  degrees of freedom and their interactions.    The  idea of Abelian gauge fixing and the resulting picture of
(Abelian) dual superconductivity mechanism  for  confinement \cite{TH}   implies 
that the most relevant  low-energy effective degrees of freedom are the  magnetic monopoles of
 two types,   carrying each unit charge with respect to  the  two $U(1)$ subgroups of   the color $SU(3)$  group.  Condensation of these
monopoles would  lead to confinement of electric charges.   This scenario, however, 
leaves  many questions unanswered.  One is the issue of chiral symmetry breaking.   Do the
Abelian  monopoles  carry flavor quantum numbers?   If so, which, and how?  Does confinement induce chiral symmetry breaking?
Also, what is  the gauge dependence of such a description?

Another, more serious  problem is this.   Does the Abelian dominance of confinement  imply dynamical color  $SU(3) \to U^2(1)$ breaking,
with a characteristic enrichment of meson  spectrum?   There are no phenomenological indication  that this  takes place in the real world
of strong interactions.    If so, what are  the other relevant   degrees of freedom, and how do they interact?  What is the structure of the 
low-energy effective action? 

Lattice QCD has not given a clear  answer  to these questions so far. 

Here we follow another approach: we examine    carefully  certain    solvable models which are basically  very  similar to QCD but  in  which
mechanism of confinement and  dynamical flavor symmetry breaking can be studied in exact, quantum mechanical 
fashon \cite{SW1}$^-$\cite{AGK}.     The models which we study with particular  attention will be  softly
broken
${ N}=2\,
$  supersymmetric gauge theories with gauge groups
$SU(n_c)$, $USp(2 n_c)$ and 
$SO(n_c)$, and with all possible numbers of fundamental quark flavors, compatible with asymptotic freedom
\cite{ArPlSei}$^-$\cite{AGK}.

\begin{figure}[ht]
\begin{center}
\leavevmode
\epsfxsize=10 cm 
\epsffile{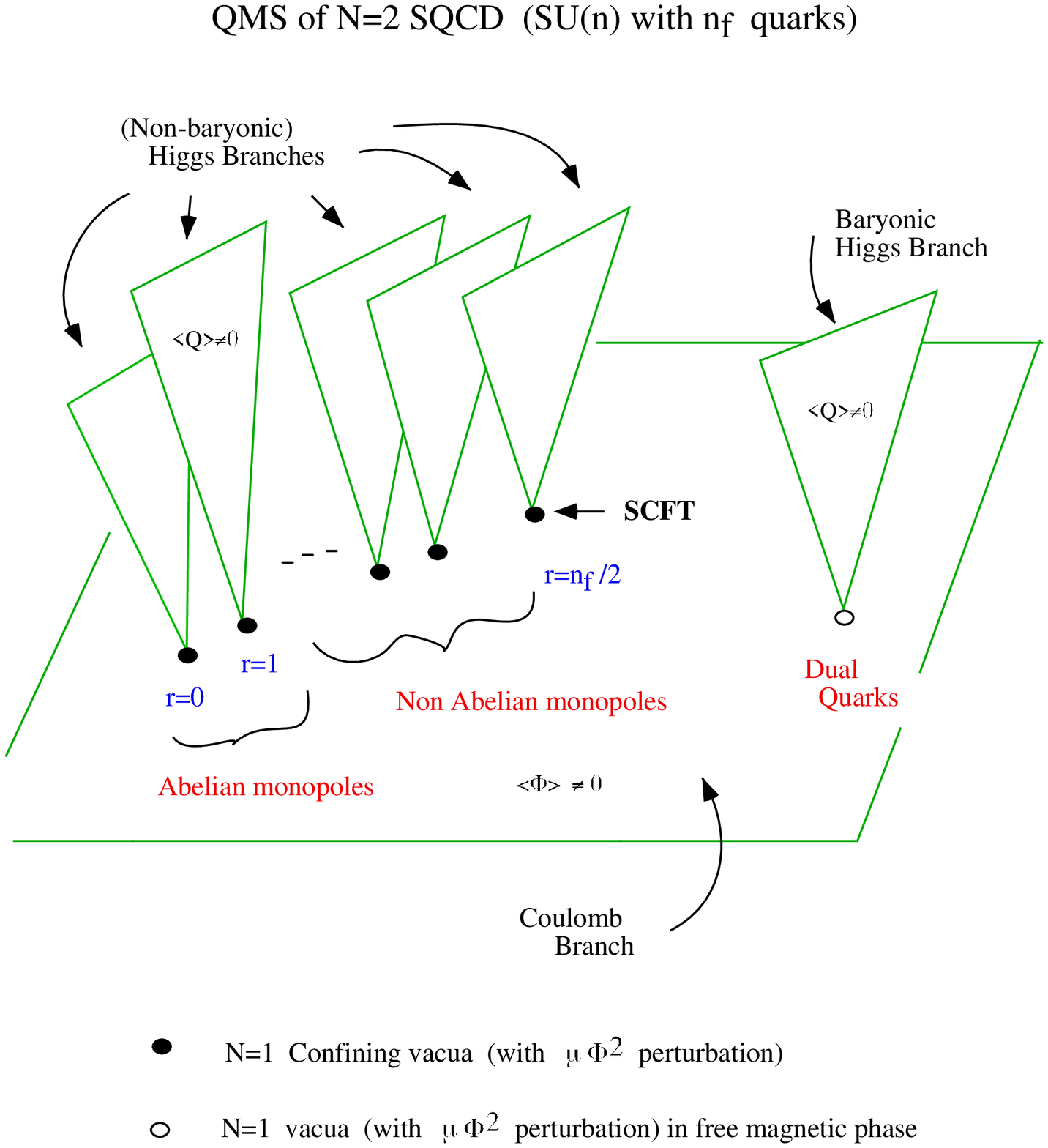}           
\end{center}
\label{fig1}
\caption{Quantum moduli space of $SU(n_c)$ theories.}
\end{figure}

\begin{figure}[ht]
\begin{center}
\leavevmode
\epsfxsize=10cm 
\epsffile{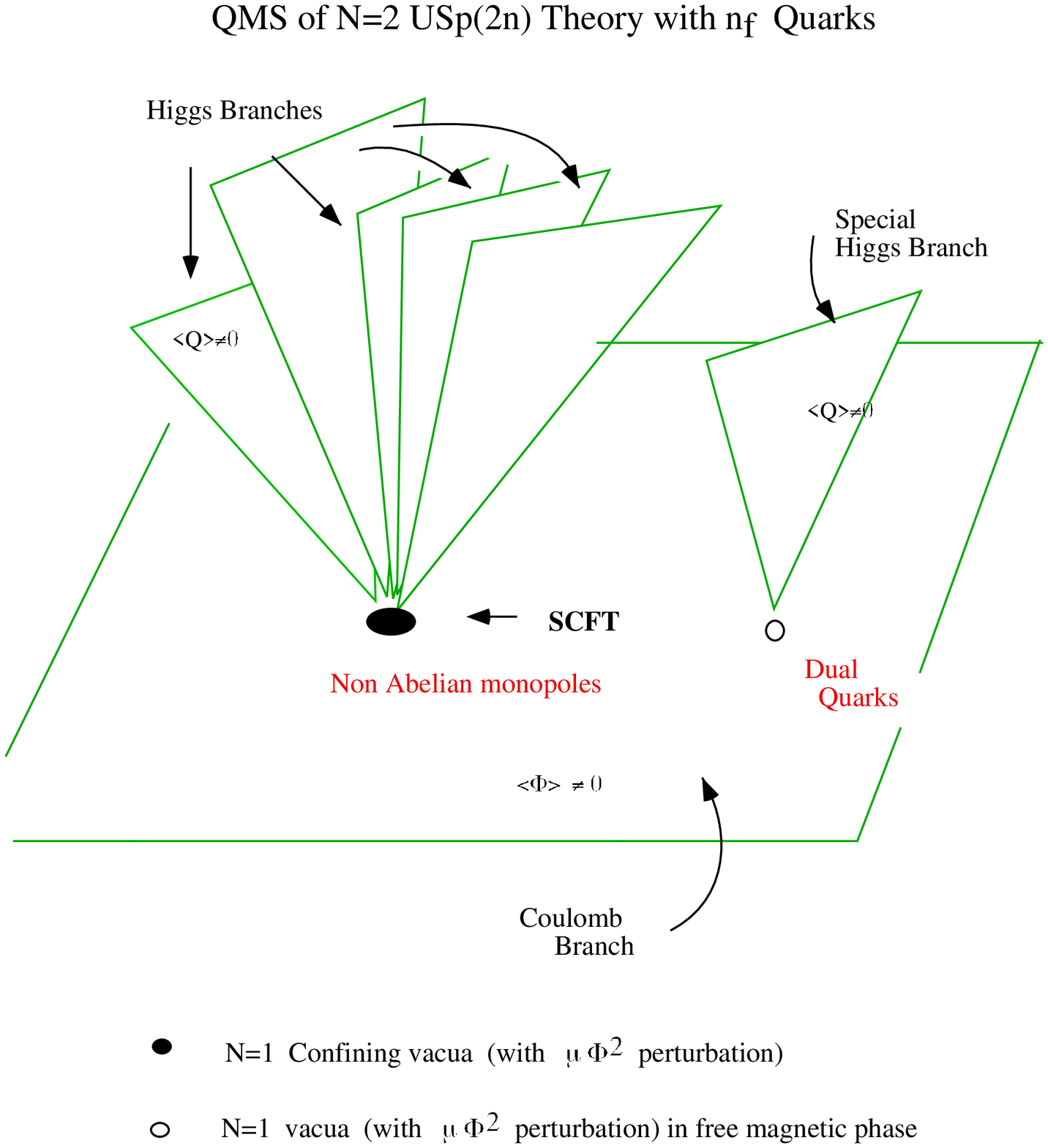}             
\end{center}
\label{fig2}
\caption{Quantum moduli space of $USp(2 n_c)$ theories.}
\end{figure}

\subsection{Models and Global Symmetry }

The Lagrangian has the structure 
\begin{equation}
{  L} =     {1\over 8 \pi} {\hbox{\rm Im}}   \, \tau_{cl} \left[\int d^4 \theta \,
\Phi^{\dagger} e^V \Phi +\int d^2 \theta\,{1\over 2} W W\right]
+ { L}^{(quarks)} + \Delta { L},
\end{equation}
where
\begin{equation} 
\Delta { L}=   \int \, d^2 \theta \, \,   \mu  \,  {\hbox {\rm Tr}}   \, \Phi^2, \qquad    
 \tau_{cl} \equiv  {\theta_0 \over \pi} + {8 \pi i \over  g_0^2} \end{equation} 
and      (${ N=}2$)  gauge multiplet
$\Phi= \phi \, + \, \sqrt2 \,
\theta  \,\psi + \, \ldots;$   $W_{\alpha} = -i \lambda \, + \, {i
\over 2} \, (\sigma^{\mu} \, {\bar \sigma}^{\nu})_{\alpha}^{\beta} \, F_{\mu \nu} \,
\theta_{\beta} + \, \ldots $   are both in the adjoint representation; 
 \begin{equation}   { L}^{(quarks)}= \sum_i \, [ \int d^4 \theta \, \{ Q_i^{\dagger} e^V
Q_i + {\tilde Q_i}^{\dagger}  e^{ {\tilde V}}    {\tilde Q}_i \} +
\int d^2 \theta 
\, \{ \sqrt{2} {\tilde Q}_i \Phi Q^i    +      m_i   {\tilde Q}_i    Q^i   \}]
\end{equation}
describes  the $n_f$  quarks and squarks. 

The number of flavor is limited to 
$$  n_f \le  2n_c, \,\, 2 n_c +2,  \,\,  n_c-2, \qquad {\hbox{\rm for}}  \quad  SU(n_c), \,\, USp(2n_c), \,\, SO(n_c),$$
 respectively,   by the requirement of   asymptotic freedom.  
 The global symmetry  of the model is    ($m_i \to 0$): 
\begin{equation}  G_F=  \cases { U(n_f) \times Z_{2n_c - n_f}  \qquad  SU(n_c); \cr
  SO(2n_f)  \times  Z_{2n_c  + 2  - n_f} \qquad  USp( 2 n_c); \cr
 USp(2n_f)  \times  Z_{2n_c   -2  n_f -4}  \qquad  SO(n_c)  } 
\end{equation}  

It turns out  that  upon  ${ N}=1$ perturbation  $\Delta { L}$,   and with  generic quark masses,  only a discrete set of 
vacua remain.  Most important of all,  vacua in confinement phase can be classified further by the type of the low-energy degrees of
freedom and by the way they interact.   See  Figs 1,2 and below.  

\subsection{   Different types of Confining Vacua   in Softly Broken 
${ N}=2$ Gauge  Theories   }    

Indeed, different types of confining vacua  are \cite{CKMP} (see also Table  \ref{tab1}): 

\begin{enumerate}

\item    
  Abelian  dual superconductor   -    with dynamical Abelianization.    
  The effective action has the form of a magnetic $U(1)^{R}$  gauge theory, where 
  $R=$   rank of $G_c$.

{\footnotesize [Examples are: $r=0,1 $ vacua in $SU(n_c);$   also   all vacua in theories with  $n_f=0$]};

\item   
  Confinement by condensation of non-Abelian dual quarks    of effective
   $SU(r)  \times   U(1)^{n_c-r+1} $  theory; 

{\footnotesize [ $r=2,3,\ldots, [{n_f-1  \over  2}]$   vacua   of $SU(n_c)$;  also 
   $USp(2n_c), SO(n_c)$ models with $m\ne 0$ ]}  
  
\item    
     Confining vacua  which are   deformed   superconformal theories  
{\footnotesize [ $r= {n_f  \over 2}$   vacua of $SU(n_c)$;  also all confining vacua 
in $USp(2n_c), SO(n_c)$ models with $m= 0$ ]. }

\item 
 There exist also   vacua in free-magnetic phase, with   no confinement, no DSB,   for theories with larger
$n_f$ (e.g.  $ n_f \ge  n_c, $   in  $SU(n_c)$.)
\end{enumerate}

We wish to find out:

\noindent  { \it   Why  does Abelianization   occur in some vacua? }

\noindent  { \it   What are the dual quarks? }  

\noindent  { \it   What degrees of freedom  are there  in SCFT   and   how do they interact? }
 \begin{table}[th]
\tbl{ Phases of $SU(n_c)$ gauge theory with $n_f$ flavors.       
$ {\tilde n}_c
\equiv n_f-n_c$.  NB  and BR   stand for the ``non-baryonic" and ``baryonic"   Higgs branches. \vspace*{1pt}}
{\footnotesize
\begin{tabular}{|c|r|r|r|r|}
\hline
{} &{} &{} &{} &{}\\[-1.5ex]
    label ($r$)     &   Deg.Freed.      &  Eff. Gauge  Group
&  Phase    &   Global Symmetry    \\[1ex]
\hline
{} &{} &{} &{} &{}\\[-1.5ex]
$0$  (NB)     &   monopoles   &     $ U(1)^{n_c-1}   $               &   Confinement
   &      $U(n_f) $      \\[1ex]
$ 1$ (NB)            &  monopoles         & $U(1)^{n_c-1} $        &
Confinement       &     $U(n_f-1) \times U(1) $    \\[1ex]
$ 2,.., [{n_f -1\over  2}] $ (NB)   &  dual quarks        &    $SU(r)
\times U(1)^{n_c-r}   $  &    Confinement
&          $U(n_f-r) \times U(r) $ \\[1ex]
$ {n_f / 2}  $ (NB)   &   rel.  nonloc.     &    -    &    Almost SCFT
&          $U({n_f / 2} ) \times U({n_f/2}) $
 \\[1ex] 
BR  &  dual quarks     &
$ SU({\tilde n}_c) \times  U(1)^{n_c -  {\tilde n}_c } $                &
Free Magnetic
&      $U(n_f) $          \\[1ex]
\hline
\end{tabular}\label{tab1} }
\end{table}

  \begin{table}[th]
\tbl{ Phases of $USp(2 n_c)$ gauge theory  with $n_f$ flavors  with
$m_i \to 0$.   $ {\tilde n}_c \equiv n_f-n_c-2$.  \vspace*{1pt}}
{\footnotesize
\begin{tabular}{|c|r|r|r|r|}
\hline
{} &{} &{} &{} &{}\\[-1.5ex]
    label ($r$)     &   Deg.Freed.      &  Eff. Gauge  Group
&  Phase    &   Global Symmetry    \\[1ex]
\hline
{} &{} &{} &{} &{}\\[-1.5ex]
1st Group  &  rel.  nonloc.       &    -    &
Almost SCFT
&          $ U(n_f)  $    \\[1ex]
2nd Group       &  dual quarks     &      $USp(2  {\tilde n}_c) \times
U(1)^{n_c -{\tilde n}_c} $               &  Free Magnetic
&      $SO(2n_f) $         \\[1ex] 
\hline
\end{tabular}\label{tab2} }
\end{table}

\section{Non-Abelian Monopoles}

\subsection{ Gauge Symmetry Breaking and Goddard-Nuyts-Olive-Weinberg monopoles}

In order to answer these questions,  let us first recall some well-known and some relatively little-known   
 facts about non-Abelian monopoles \cite{GNO,BK}.    The relevant setting is a gauge theory in which gauge symmetry is broken
spontaneously as  
\begin{equation}      G   \,\,\,{\stackrel {\langle \phi \rangle
     \ne 0} {\longrightarrow}}     \,\,\, H  \end{equation} 
where $H$ is in general non-Abelian.  Finite energy classical configurations are such that
$$  { D} \phi    \,\,\,{\stackrel {r \to   \infty  } {\longrightarrow}}   \,\,\,0    ,   \quad    \Rightarrow   \quad 
\phi \sim   U \cdot  \langle \phi \rangle  \cdot U^{-1}  \sim       \Pi_2(G/H)  =
\Pi_1(H)
$$  
\begin{equation}  A_i^a  \sim  U \cdot {\partial_i  }  U^{\dagger}   \to      \epsilon_{aij}  { r_j  \over     r^3}   G(r): 
 \end{equation}  
they represent elements of the homotopy group $ \Pi_1(H). $
Asymptotically we can take
\begin{equation}     G(r)  =    \beta_i   T_i,   \qquad   T_i  \in  {\hbox {\rm  Cartan S.A.  of }} \, \,\,  H  
 \end{equation}   
so that the constant vectors $ \beta_i $ characterize the configurations.

Topological quantization leads to the result that 
$$   \beta_i  =   {\hbox {\rm  weight vectors  of}} \,\,  {\tilde H} =  {\hbox {\rm  dual   of}} \,\, H,   
$$
where examples of duals of gauge groups are:
$$      {\tilde H}     \Leftrightarrow   H  $$ 
{\large 
\begin{table}[h]   
\begin{center}
\begin{tabular}{c  c   c}
\hline  
$SU(N)/Z_N       $        &   $\Leftrightarrow$                 &    $SU(N)     $          \\
  $ SO(2N)  $     &   $\Leftrightarrow$    &   $SO(2N) $       \\  
  $ SO(2N+1)  $     &   $\Leftrightarrow$     &   $USp(2N) $       \\ \hline
\end{tabular}
\label{tabtheta}
\end{center}
\end{table}} 
Note that as  $|\phi| \to \infty$ these finite energy solutions become singular 
  Dirac type monopoles.   Also, in the simplest case of    $G=SU(2)$,  $H=U(1)$
they reduce to the well known 
 't Hooft-Polyakov  monopoles.

\subsection{  Quantum Numbers of N.A. monopoles}    

In order to see what quantum numbers these monopoles carry, let us consider first the simplest case
$$     SU(3) {\stackrel {\langle \phi \rangle } {\longrightarrow}}     SU(2) \times  U(1), \qquad   \langle \phi\rangle = 
 \pmatrix{  v & 0& 0  \cr  0 & v & 0 \cr  0&0& -2v  } 
$$
Consider  the  subgroup $SU_U(2)  \subset SU(3) $
{\footnotesize    $$   t^4= { 1\over 2}   \pmatrix{  0 & 0& 1  \cr  0 & 0 & 0 \cr  1  &0& 0   }; \quad  
t^5= { 1\over  2}   \pmatrix{  0 & 0& -i  \cr  0 & 0  & 0 \cr i &0& 0    }; \quad  
{ t^3 +  \sqrt3 t^8 \over 2}  = { 1\over 2}   \pmatrix{  1  & 0  & 0  \cr  0 & 0 & 0 \cr  0&0& -1   }; \quad  
$$     }    
which is broken as 
$$       SU_U(2) {\stackrel {\langle \phi \rangle } {\longrightarrow}}     U_U(1).  $$
Use 't Hooft-Polyakov  solution   for  $ \phi(r), A(r)$   for the broken $SU_U(2)$,     one finds a   $SU(3)$ solution  (Sol. 1) : 
\[ \phi=
   \left( \begin{array}{ccc}
     -\frac{1}{2}v&0&0\\
     0&v&0\\
     0&0&-\frac{1}{2}v\\
   \end{array} \right)   +
   \frac{3}{2} v \Big( t_4,t_5,\frac{t_3} {2} + \frac{\sqrt{3} t_8}{2} \Big)
   \cdot \hat{r} \phi(r), \]
\begin{equation} \vec{A}=  \Big( t_4,t_5,\frac{t_3} {2} + \frac{\sqrt{3} t_8}{2} \Big)
   \wedge \hat{r} A(r).      \qquad  \end{equation}
\noindent  Another solution  (Sol. 2)  can be found by considering another      $SU_V(2)  \subset SU(3) $
{\footnotesize $$   t^6= { 1\over 2}   \pmatrix{  0 & 0&  0  \cr  0 & 0 & 1 \cr  0  & 1 & 0   }; \quad  
t^7= { 1\over 2}   \pmatrix{  0 & 0& 0  \cr  0 & 0  & -i  \cr 0 & i & 0    }; \quad  
{ -  t^3 +  \sqrt3 t^8 \over 2}  = { 1\over 2}   \pmatrix{  0  & 0  & 0  \cr  0 & 1 & 0 \cr  0&0& -1   }, \quad  
$$  } 
leading  to a degenerate doublet of monopoles with  charges.    
{       
\begin{table}[h]      
\begin{center}
\begin{tabular}{c  c   c}
 \\
  monopoles    &    ${\tilde  {SU}  (2)  } $           &    ${\tilde U(1) }     $          \\    \hline 
  ${\tilde q}  $        &     $ {\underline 2 }$     &      $1$       \\  \hline 
\end{tabular}
\label{tabtheta}
\end{center}
\end{table}   }

\subsection{  Generalization}  
Generalization to the case of the symmetry breaking
$$     SU(n) {\stackrel {\langle \phi \rangle } {\longrightarrow}}     SU(r) \times  U^{n-r}(1), \qquad   \langle \phi\rangle = 
 \pmatrix{  v_1  {\bf 1}_{r\times r}   & {\bf 0 } &  \ldots &   {\bf 0}   \cr  {\bf 0 }  & v_2   & 0   & \ldots   \cr  {\bf 0}  &0& \ddots 
&  
\ldots
\cr {\bf 0}    & 0   &  \ldots  & v_{n-r+1}   }   
$$
can be done by considering  various   $SU_i(2)$  subgroups  ($ i=1,2,\ldots, r$)   living in $[  i , r+1  ]  $  subspace: one finds
(see the Table below)  
  \begin{romanlist}

\item       Degenerate  $r$-plet of monopoles  ($ q$);  
  
\item    Also,      Abelian monopoles  ($e_i$),   ($i=1,2,\ldots,  n-r-1$)     of     $U^{n-r-1}(1)$  (non degenerate) appear;

 \begin{table}[h]      
\begin{center}
\begin{tabular}{c  c   c c c  c c  }
 \\
  monopoles    &    ${\tilde  {SU}}  (r)   $           &    ${\tilde U_0(1) }     $      &    ${\tilde U_1(1) }     $     &    ${\tilde U_2(1) }  $   &
$\ldots  $  &         ${\tilde U_{ n-r-1} (1) }  $        
\\    \hline 
  $ q  $        &     $ {\underline r   }$     &      $1$     &  $0$ &   $0$ & \ldots   &  $0$    \\  \hline
 ${ e}_1  $        &     $ {\underline 1 }$     &     
$0 $     &   $1$  &  $0$     & $\ldots $     &  $0$    \\  \hline 
${ e}_2 $        &     $ {\underline 1  }$     &   $0 $     & $0 $  &  $1$   & $0$    &  $0$  \\  \hline 
$\vdots $        &     $ {\underline 1   }$     &      $0 $     &  $ \ldots $    &   &   &   $0$    \\  \hline 
${ e}_{n-r-1}    $        &     $ {\underline 1  }$     &      $0$     & $0$     &  $ \ldots$   &  $\ldots $     &  $1$   \\  \hline 
\end{tabular}
\label{tabtheta}
\end{center}
\end{table}

\item  These monopoles have  {\it the same charge structures found in the  $\, r\,$-vacua of  $ {  N}=2$  SQCD }   (!) 

\item  Also,  the flavor quantum numbers   of     non-Abelian   monopoles can be understood by the generalized  Jackiw-Rebbi
mechanism\cite{BK}. 
 
\end{romanlist}

\subsection{   Subtleties}    

There are certain  subtleties around  the non-Abelian monopoles:  

\begin{romanlist}

\item   ``Colored dyons" have been shown not to exist.\cite{CDyons}  
{\it Actually there  is no paradox here.}    Non-Abelian  monopoles carry both Abelian and non-Abelian  charges, but both refer to   ${\tilde
H},$  not $H$ itself, while the results of Abouelsaood et.al. \cite{CDyons}  refer to a non-Abelian generalization of charge fractionalization,
which is not possible;  

\item   Non-Abelian monopoles are to transform as members of various  multiplets of the dual group ${\tilde H}$, {\it not }  of  $H$
itself.  Any search for the ``gauge  zero modes"  should involve  {\it non-local field transformations};

\item  It is {\it not }   justified to study the system
$$   G   \,\,\,{\stackrel {\langle \phi \rangle    \ne 0} {\longrightarrow}}     \,\,\, H  $$ 
as a limit of maximally broken cases   ($H_0 \subset$ Cartan S.A.  of $G$):
$$  \langle \phi \rangle =  {\bf h} \cdot {\bf  H_0}, \qquad  h_i \to 0, \quad {\hbox {\rm for}}\quad  H_{0i}
\subset  H. $$
To do so would necessarily lead one to  the (non-semi-classical)  domain of strongly coupled, infinitely extended, light
monopoles     (just think of taking the limit   $v \to 0 \,\,$  to   study   the 't Hooft-Polyakov monopole of   $\,\,\,SU(2)   \,\,\,{\stackrel v
{\longrightarrow}}    
\,\,\, U(1)
$  theory!).

\item  Indeed, non-Abelian monopoles  are never really  semi-classical, even when 
 $$\langle \phi \rangle  \gg \Lambda_H, $$      
if   $H$ interactions   grow   strong in   the IR:    $H$    may be   further  {\it dynamically}    broken  at $\mu
\sim
\Lambda_H$.     
If it  is, ``non-Abelian monopoles" simply means a set  of       {\it approximately}    degenerate  monopoles
\footnote { We verified this   explicitly by using the  formula of Klemm et. al. \cite{curves} 	in  ${ N}=2$ susy  $SU(3)$ pure
Yang-Mills theory   in an appropriate region of quantum moduli space.}.

\item Only if  $H$ remains   unbroken do   non-Abelian monopoles in an irreducible representation  of  ${\tilde
H}$  make appearance in the low-energy action.

\item  Most remarkably,    
this last option  seems to be realized in the  $r$-vacua  of $SU(n_c) $, $n_f$ theories. We propose that   {\it  the dual quarks are nothing but
the        GNO monopoles.   }

\end{romanlist}

\subsection {   Duality }    

Further justification of our ideas  comes from the duality considerations. 

\begin{itemize} 

\item      $r$ vacua with 
   $SU(r)  \times   U(1)^{n_c-r+1} $  gauge group  occur     only   for
$$   r  <   {n_f \over  2  }.$$
This can be understood as due to the      sign-flip of the  beta  function:    
\begin{equation}   b_0^{(dual)} \propto   -  2 \, r  +   n_f  >  0,   
\qquad       b_{0} \propto  -  2 \, n_c +    n_f  <     0,      
\label{betafund}  \end{equation} 
so that the low energy $SU(r)$ interactions are infrared-free.  Note that for this to happen 
the flavor-dressing of the monopoles  is   essential.

\item   When this sign flip is  not possible for some reason,  such as in   pure ${ N}=2$ YM or in   generic points of QMS  of ${
N}=2$  theories,       dynamical Abelianization occurs.

\item   These questions are  related to the  resolution of the old  Dirac-quantization-vs-Renormalization-Group  puzzle   (i.e., why the 
quantization condition 
 $$  g_e (\mu)  \cdot g_m(\mu) = 2 \pi n, \qquad \forall \mu   $$
is valid at any scale $\mu$?) 
in the Seiberg-Witten model.

\item The boundary,     $r= {n_f \over 2}$ case,     is a    SCFT  (nontrivial  IR  fixed point):
    non-Abelian monopoles and dyons   still  show up as recognizable low-energy effective degrees of freedom,  although  their interactions are
nonlocal. 

\end{itemize}

\begin{figure}[ht]
\begin{center}
\leavevmode
\epsfxsize  5   cm    
\epsffile{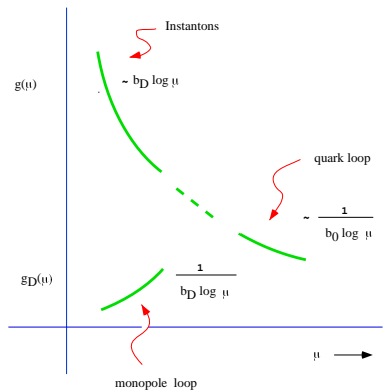}           
\end{center}
\caption{Duality. Monopole loop is equivalent to infinite instanton sum. } 
\label{duality}
\end{figure}

\subsection {     Dynamical Symmetry Breaking: a Puzzle
 }

\begin{figure}[ht]
\begin{center}
\leavevmode
\epsfxsize=8  cm 
\epsffile{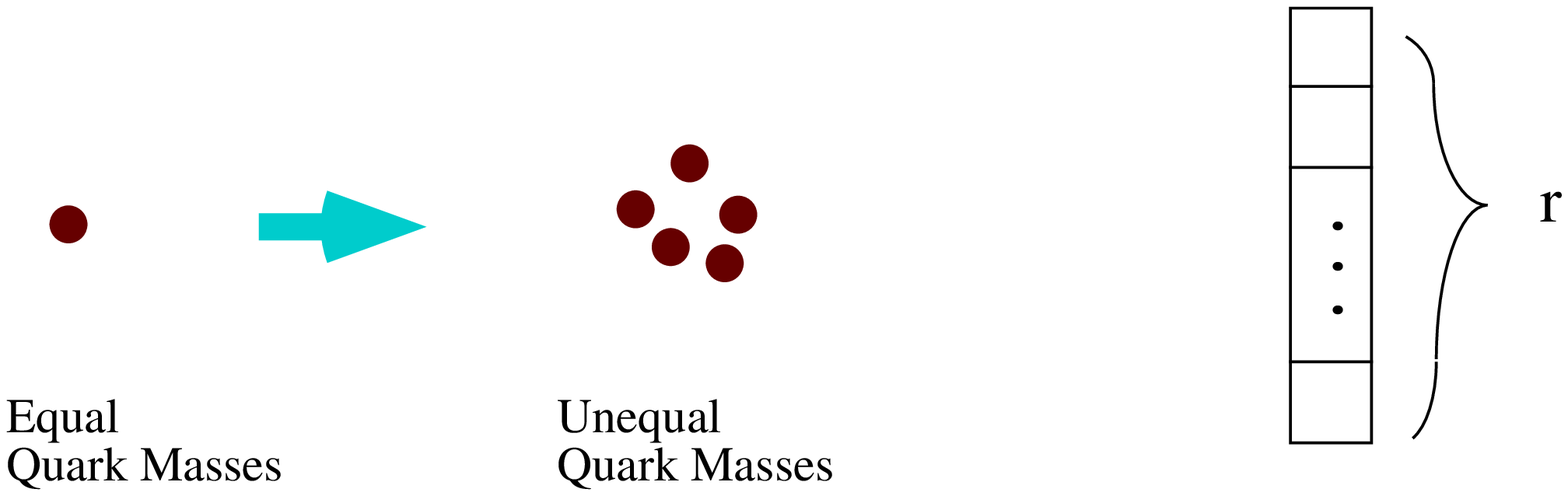}             
\end{center}
\caption{}
\end{figure}  
  
\begin{itemize} 

\item As the quark masses are chosen unequal, $m_i\ne m_j$, each of $r$  vacua  splits into  $  {n_f \choose  r}$   points in QMS.   
This is very suggestive of a possibility that the massless monopoles in each vacuum is an   (Abelian)  monopole in ${\underline  {n_f
\choose  r}}$  representation   of the global   
$SU(n_f)$. This is precisely what happens in  the  $SU(2)$  theory with  $n_f=1,2,3$. 
This would however (for generic $SU(n_c)$  theories)  lead  to an effective action 
  with an  accidental   global  $SU({n_f \choose  r})$ symmetry and hence to an enormous  number of Nambu-Golstone bosons   
when these field condense. 

\item   Actually this does not happen. The system avoids this  awkward  situation by having 
  non-Abelian  monopoles in ${\underline  {  r}}$   of  dual  color $SU(r)$, and  in the fundamental representation   ${\underline  { n_f
}}$   of  the   global 
$SU(n_f)$.    They      condense \cite{CKMP}     in color-flavor diagonal fashion
\begin{equation}   \langle  q_{\alpha }^i \rangle    =   \delta_{\alpha }^i \,  v, \qquad   \alpha =1,2,\ldots, r, \quad i=1,2,\ldots n_f
\end{equation}    
(``Color-Flavor-Locking"), breaking the global symmetry as
\begin{equation}  G_F = SU(N_f) \times U(1)  \Rightarrow   U(r) \times U(n_f-r).  \label {symbr} \end{equation}

\item  The non-Abelian monopoles may be regarded as baryonic constituents of  the Abelian monopole, 
$$   U(1) \, {\hbox{\rm monopole}}   \sim   \epsilon^{a_1   \ldots  a_r} {q_{a_1}^{i_1}  q_{a_2}^{i_2}
\ldots q_{a_r}^{i_r} }.     $$
The Abelian monopole,   $SU(r)$ being  infrared free, breaks up into the former!

\end{itemize}

\section{     Almost Superconformal   Confining Vacua    }

The most interesting sort of confining vacua  we encounter in the softly broken ${ N}=2$  
supersymmetric gauge theories are however  those which appear as deformation (perturbation) 
of a  nontrivial superconformal theory \cite{AD,SCF,Eguchi}.   In order to be concrete, let us  study the case of the sextet vacua in 
$SU(3)$, $n_f=4$  (${ N}=2$)    supersymmetric QCD in some detail below \cite{AGK}.   

\subsection{   Sextet Vacua of  $SU(3)$, $n_f=4$ Model}  

The Seiberg-Witten  curve  of this theory {\it equal bare quark masses}    ($m_a=m$)  is \cite{curves}    
$$  y^2 =  \prod_{i=1}^{3}  (x-\phi_i)^2 -    (x+m )^4    \equiv    (x^3  -   U  x  -
V )^2  -  (x+m )^4.  
\label{SW} $$
At the sextet vacua  of our interest ($diag\,  \phi =  (-m, -m,  2m )$),
$    U=   \langle {\hbox{\rm Tr }}     \Phi^2   \rangle = { 3 \, m^2},\,\,$   $ V= \langle {\hbox{\rm Tr }}  \Phi^3   \rangle  =    2 \, m^3,$   the curve
exhibits  a singular behavior,
$   y^2 \propto   (x+m)^4
$ 
corresponding to the unbroken $SU(2)$  symmetry.

The well known  mass formula  is  
$$  M_{ (g_{1}, g_{2};  q_{1}, q_{2})} =
\sqrt 2 \,  | g_{1}\, a_{D1} +  g_{2}\,a_{D2}  +    q_{1}\,a_{1} +
q_{2} \, a_2 |,        $$
$$    a_{D1}  =  \oint_{\alpha_1}  \lambda, \qquad  a_{D2}  =
\oint_{\alpha_2}  \lambda, \qquad
 a_{1}  =  \oint_{\beta_1}  \lambda, \qquad  a_{2}  =  \oint_{\beta_2}  \lambda,
$$ where the (meromorphic)  one-form $\lambda$ is given by
$$
\lambda =  { x \over 2 \pi}  d\, \log{  \prod (x -\phi_i) -y \over  \prod (x -\phi_i)+y  }. 
 $$ 

\subsection{Expansion near the SCFT Point }  

\begin{figure}[ht]
\begin{center}
\leavevmode
\epsfxsize=6 cm  
\epsffile{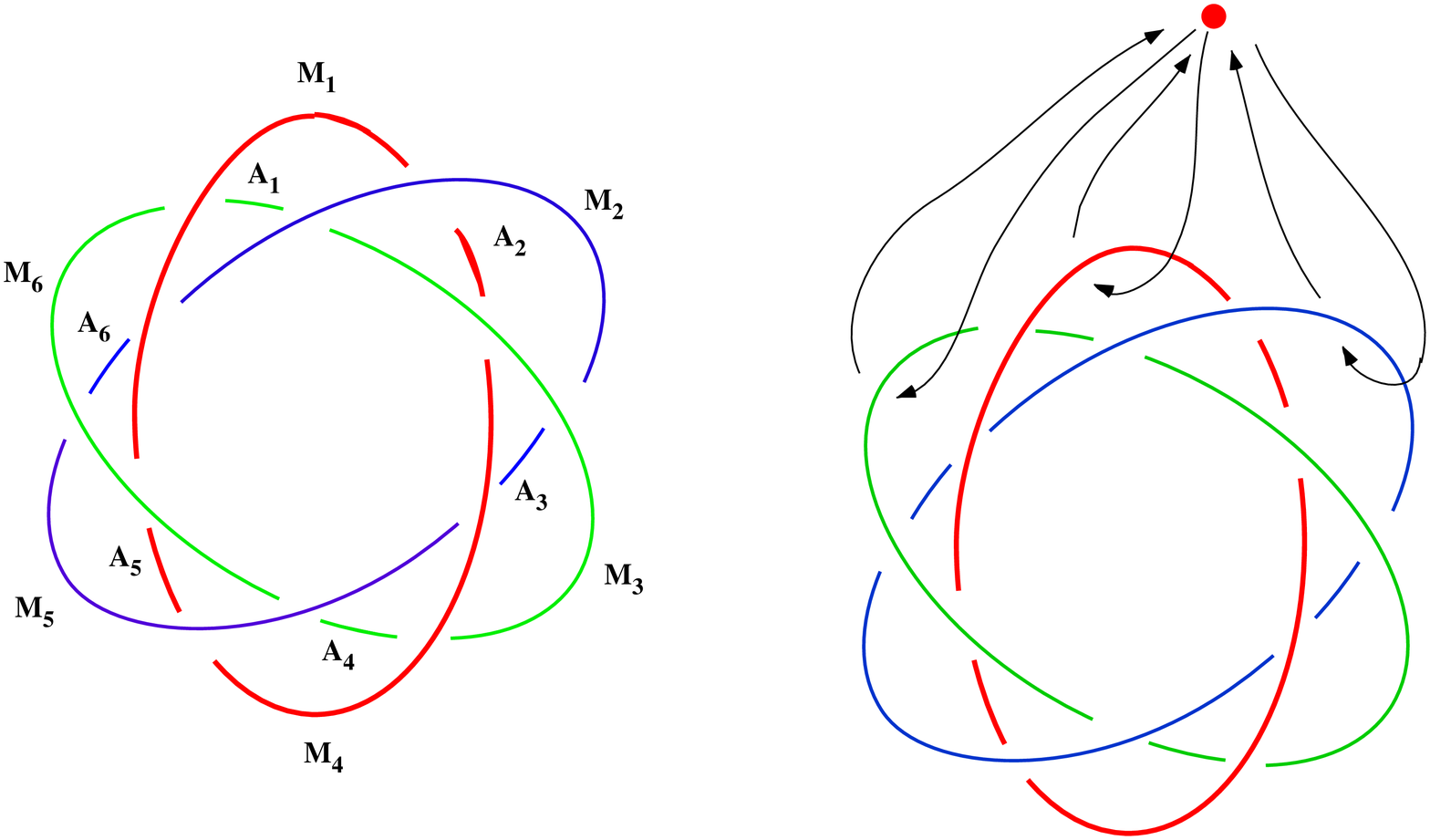}               
\end{center}
\caption{Loci in $u,v$ space where certain particles become massless. }
\label{Rings}     
\end{figure}

In order to find out the nature of the low-energy massless fields present, one has to expand around the singularity, 
$$  U=    { 3 m^2} +u, \qquad V=  2 m^3 + v.
$$
The discriminant of the curve factorizes as \cite{Eguchi} 
$$   \Delta=  \Delta_s \, \Delta_{+} \,  \Delta_{-},\qquad   \Delta_s =  (m \, u - v)^4
$$ 
so   the loci of  $\Delta=0$  are
$$ v=m\,u, \qquad  v=  m\,u +    \frac{u^2}{4}, \qquad v=  m\,u -
\frac{u^2}{4}. 
$$
By rescaling  ~~
 $u =  m \, {\tilde u}, \,\,   v=   m^2 \, {\tilde v }$,   and   intersecting  them with   a   $S^3$   
\[  |{\tilde u}|^2 + |{\tilde v}|^2 =1. \]   and making  a  stereographic projection  from  $S^3  \to
R^3$, one finds  that   the curves (in $u,v$  space)  along which some particles become massless take the form of the 
three linked rings (Fig. \ref{Rings}).     
       
\subsection  {Monodromy and Charges}

In order to find what  charges are carried by these massless particles, one has then to  study 
the monodromy  transformations (among    $a_{D1},\,a_{D2},\, a_{1},\, a_2$)  as one moves along  various closed curves 
encircling  parts of the linked rings. 
{
 \begin{figure}[ht]
\begin{center}
\leavevmode
\epsfxsize=6 cm  
\epsffile{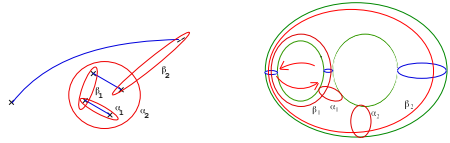}                 
\end{center}
\caption{Homology cycles and the transformation exchanging the two necks of the bitorus.}
\label{twonecks}
\end{figure}
   }
For instance  the monodromy around  $M_1$   leads to  
\[  \alpha_1 \to \alpha_1, \qquad  \beta_1 \to   \beta_1 - 4 \alpha_1, \qquad
\alpha_2 \to \alpha_2, \qquad  \beta_2 \to \beta_2.
\]
namely,
{\small 
\begin{equation}    \pmatrix{a_{D1}  \cr   a_{D2}  \cr  a_1 \cr  a_2 }   \to  M_1
\pmatrix{a_{D1}  \cr   a_{D2}  \cr  a_1 \cr  a_2 },
\qquad
  M_1= {\tilde M}_1^4, \quad  \, {\tilde M}_1=\pmatrix   { 1 & 0& 0  & 0 \cr
0& 1 & 0& 0 \cr   -1 & 0& 1& 0\cr
 0& 0& 0&  1  }   \label{M1}  \end{equation}   From  the  formula 
\begin{equation}   M=  \pmatrix  {{\bf 1} +  {\vec q} \otimes  {\vec g} &   {\vec q} \otimes
{\vec q}  \cr
  - {\vec g} \otimes  {\vec g}  &     {\bf 1} -    {\vec g} \otimes  {\vec q} }
\label{charges}  \end{equation}  
 }
\noindent the (four)  massless particles   at the singularity ${\tilde v}={\tilde u}$ 
are   found to  carry  charges
$$    (g_{1}, g_{2}; q_{1}, q_{2})  =   (1,0;0,0),  $$
i.e.,   they are  four magnetic monopoles carrying the unit charge with respect to the first $U(1)$.  
Analogously:
{\small   
\[ M_{2}  =  \pmatrix   { -1 & 0& 1  & 0 \cr    0& 1 & 0& 0 \cr   -4 & 0& 3& 0\cr
 0& 0& 0&  1  }, \qquad  M_{6}  =  \pmatrix   {1 & 1& 1  & 0 \cr    0& 1 & 0& 0
\cr   0 & 0& 1& 0\cr
 0& -1& -1 &  1  }, \quad  etc. \] }

By using  then the {\it  conjugation}  relations  among   the monodromy matrices 
{\small     
\begin{eqnarray}   && M_1=  M_6^{-1}   A_5  M_6, \quad     A_2 =  M_2^{-1}  M_1  M_2, \quad
M_4 =  M_3^{-1}   A_2  M_3, \quad   A_5 =M_5^{-1}   M_4  M_5,
\nonumber\\  &&  M_2 = M_1^{-1}   A_6  M_1, \quad     A_3 =  M_3^{-1}  M_2  M_3,
\quad   M_5 =  M_4^{-1}   A_3  M_4, \quad   A_6=  M_6^{-1}   M_5  M_6,
\nonumber\\  &&  M_3 = M_2^{-1}   A_1  M_2, \quad     A_4 =  M_4^{-1}  M_3  M_4,
\quad   M_6 =  M_5^{-1}   A_4  M_5, \quad   A_1 =  M_1^{-1}   M_6  M_1
\nonumber\end{eqnarray}  }
they can be uniquely determined.  They correspond to the 
charges  
{\small 
\begin{eqnarray}  && M_1: \, (1,0;0,0)^4, \quad M_4:\,\, (-1,1; 0,0)^4, \quad  M_2: \, (-2,0;1,0), \quad M_5:\,\, (2,-2; -1,0), 
\nonumber\\  &&  A_2:\,\,  (-1, 0; 1,0)^4, \quad  A_5:\,\, (1,-1; -1,0)^4,   \quad A_3:\,\, (-
2, 2; -1,0), \quad  A_6:\,\, (2,0; 1,0),
\nonumber\\ && M_3: \, (0,1;-1,0), \quad M_6:\,\, (0,1; 1,0), \quad A_4:\,\, (4, -3;
-1,0), \quad  A_1:\,\, (-4,1; 1,0).   
\nonumber\end{eqnarray}     }  
Now  
\begin{enumerate}
\item      {\it How are  these $U(1)$ charges   related to   $SU(2) \times U(1)$ ?}
\item   {\it   Which of them are actually  there at the SCFT Point? }
\item    {\it   How do they give   $\beta=0$ ?  }
\item    {\it    How do they interact  ?     }  
\end{enumerate}

The first of these questions can be answered by studying the effect of transformation
which exchanges the two necks of the bi-torus (Fig. \ref{twonecks}),  
$$  \alpha_1 \to \alpha_2 - \alpha_1;  \qquad \beta_1 \to - \beta_1; \qquad
 \alpha_2 \to  \alpha_2;\qquad  \beta_2 \to \beta_1 + \beta_2.
$$
This allows us  to introduce a new basis such that one of the $U(1)$ factors is  a subgroup of $SU(2)$  
and  another   is 
orthogonal to it.  In the new basis,  the charges look as in Table \ref{monopcharges}.   
{\footnotesize    
\begin{table}[h]       
\begin{center}    
 \begin{tabular}  {|l|l|} \hline Matrix & Charge \\
\hline
$M_1,M_4$ & $(\pm 1,1,0,0)^4$  \\ $A_2, A_5$ &$(\pm 1,-1,\mp 1,0)^4$ \\
$M_2,M_5$ &$(\pm 2,2,\mp 1,0)$ \\ $A_3,A_6$ & $(\pm 2,-2,\pm 1,0)$  \\
$M_3,M_6$&$(0,2,\pm 1,0)$  \\ $A_1,A_4$ & $(\pm 4,-2,\mp 1,0)$  \\
\hline
\end{tabular}   
\end{center}
\label{monopcharges}      
\end{table}
}       

 \subsection{   Superconformal Limit ( $u=v=0$ )  }  

We must  first of all  {\it define  } SCFT  limit appropriately.  
 
\begin{figure}[ht]
\begin{center}
\leavevmode
\epsfxsize=8 cm    
\epsffile{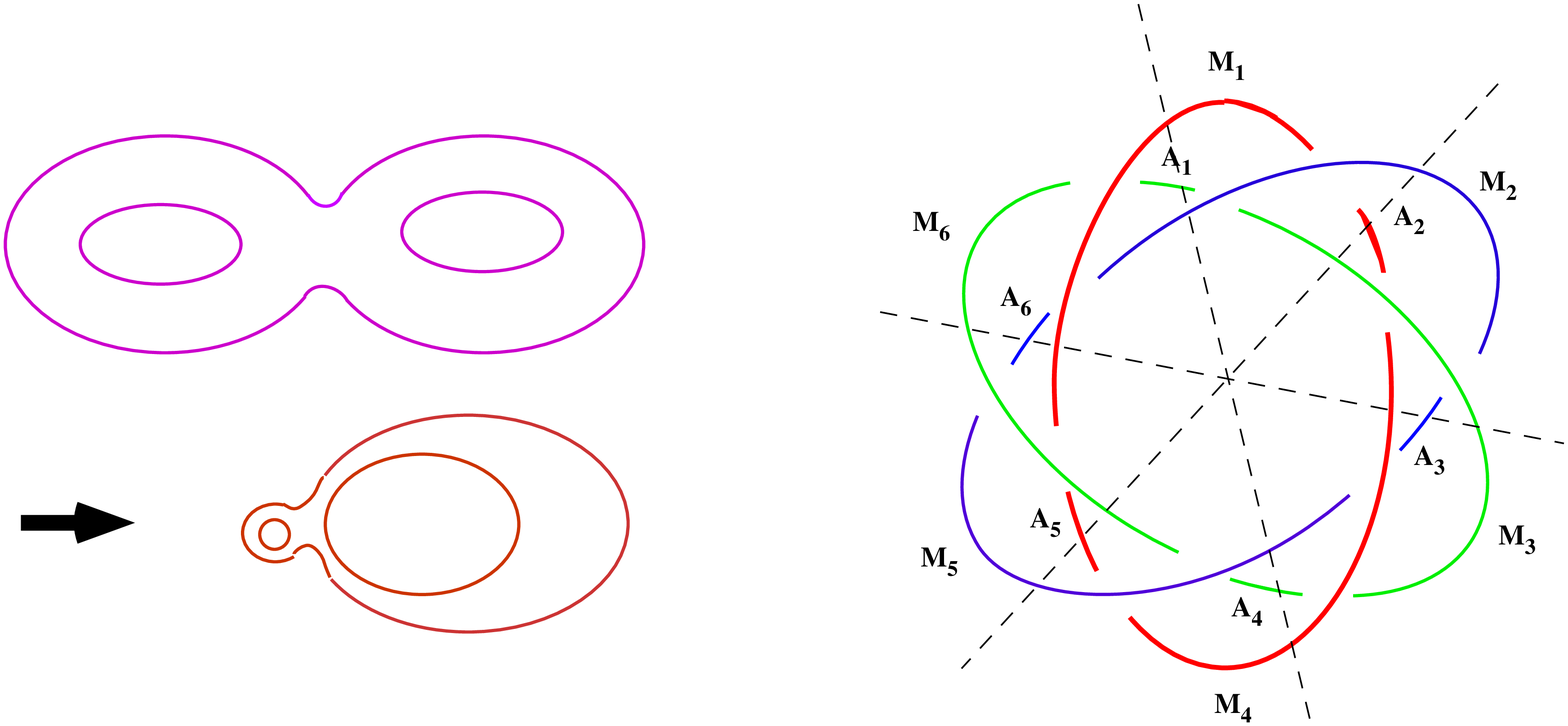}                      
\end{center}
\label{susy}
\end{figure}

  \begin{itemize}

\item  As $u, v \to 0$,  the bitorus degenerates.   
 If the branch points   $\{ b,c,d,e,f,g  \} $  collapse to  $  ( a,a,a, 0,1,\infty) $ then  $\tau$  becomes diagonal with 
(Lebowitz \cite{lebowitz})
 $$
a={{ {\vartheta}_3}^4(0  \, |  \tau_{22} )
 \over{{ \vartheta}_2}^4(0\, | \tau_{22}   ) }; \qquad   
\frac{c-b}{d-
b}= { {{\vartheta}_3}^4(0  \, |  \tau_{11} )
 \over{{ \vartheta}_2}^4(0\, | \tau_{11}   ) }.$$

\item   In our sextet vacua the curve has the singular  form 
 $$ y^2=(x^3-u x -v)^2-x^4\to    x^4   (x+1) (x-1). $$
By an apporpriate change of the variable $x$,  one finds for 
the modular parameter   of the large  torus:   $\tau_{22} \to 1  $  (weakly interacting $U(1)$ theory);

\item  As for  the small torus  ( $SU(2)$ ),       $\tau_{11}$  apparently  depends on the way  $u,\, v $ are taken to $0$.  

\item  By studying the simplified  curve  $ (x^2-u x -v)(x^2+u x +v)=0 $  with  variable change, 
$$  v=\epsilon^2; \qquad    u=\epsilon \rho;
\qquad     x=\epsilon z; \qquad    y=\epsilon^2 w,   $$ 
one  finds that    $\tau_{11}$  depends only  on  $\rho.$  
\item In other words,   different sections  of the   linked rings  at different phase of  $\epsilon $   
are  different  ($SU(2,Z)$-related)    descriptions of the same physics!   

\end{itemize}

Thus     we  {\it define}    the  SCFT  by taking the limit  
$ \quad \epsilon \to 0,  \quad     \rho \to
0,       $
namely, $\rho \to 0$ first.    This finally yields the following  charges of the massless particles in different sections.   Note the three-fold
periodicity. 

{\footnotesize     
\begin{table}[h]
\begin{center} 
\begin{tabular}{|l|l|l|l|l|l|l|l|}
 \hline $+2 i$  &$(0,1) $ &$(4,-1)$
&$(0,1)$ &
$(0,-1) $ &$(-4,1)$ & $(0,-1)$ &  $  \ldots $    \\
 \hline $ 2$ &$(2,1)$ &$(2,-1) $
&$(2,-1)$ &$(-2,-1)$ &
$(-2, 1)$ &$(-2, 1)$ & $\ldots $  \\
\hline $-2 i$  &$(0,-1) $ &$(-4, 1)$
&$(0,-1)$ &
$(0,1) $ &$(4,-1)$ &  $(0,  1) $ &   $  \ldots $  \\
\hline $-2$ &$(-2,-1)$ &$(-2, 1)$
&$(-2, 1)$ &$(2,1)$ &
$(2,-1)$ & $(2,  - 1)$ &  $\ldots $  \\
\hline $\infty$ &$(\pm 1,0)^4$ &$(\mp 1,0)^4$
&$(\pm 1,\mp 1)^4$ &
$(\mp 1,0)^4$ &$(\pm 1,0)^4$& $(\mp 1, \pm 1)^4$  
&  $\ldots $ 
\\
\hline
\end{tabular}
\end{center}
\end{table}           
}     

\begin{enumerate}
\item  The three sections are related   by unimodular transformations
2  $$  p_1=\pmatrix{-1& 4 \cr 0 & -1 };\qquad 
p_2=\pmatrix{-1& -4 \cr 1 & 3 }; \qquad   p_3=\pmatrix{1& 0 \cr 1 &
1}.  $$  
\item At $\rho=0$,  the small  branch points are  at  $(2 i,\,-2 i,\,2,\,-2)$
so that one finds  for $\tau_{11}$
$$  { 1\over 2 }  =    {  {{\vartheta}_3}^4(0  \, |  \tau_{11} )
 \over   {{\vartheta}_2}^4(0\, | \tau_{11} )  }, \label{modular} $$
which has solutions     
$$ \tau_{11}=\frac{\pm 1+i}{2}, \qquad  \frac{\pm 3+i}{10},   \qquad   \ldots \label{tau}  $$
{\small    Other solutions by $SL(2, Z)$  transformations   $ \tau \to  \tau+2; \,\, 
  \tau \to  {\tau  \over   1-2 \tau  }. $}
\end{enumerate}

\subsection{     Renormalization-Group  Fixed Point }     

\begin{figure}[ht]
\begin{center}
\leavevmode
\epsfxsize=8   cm     
\epsffile{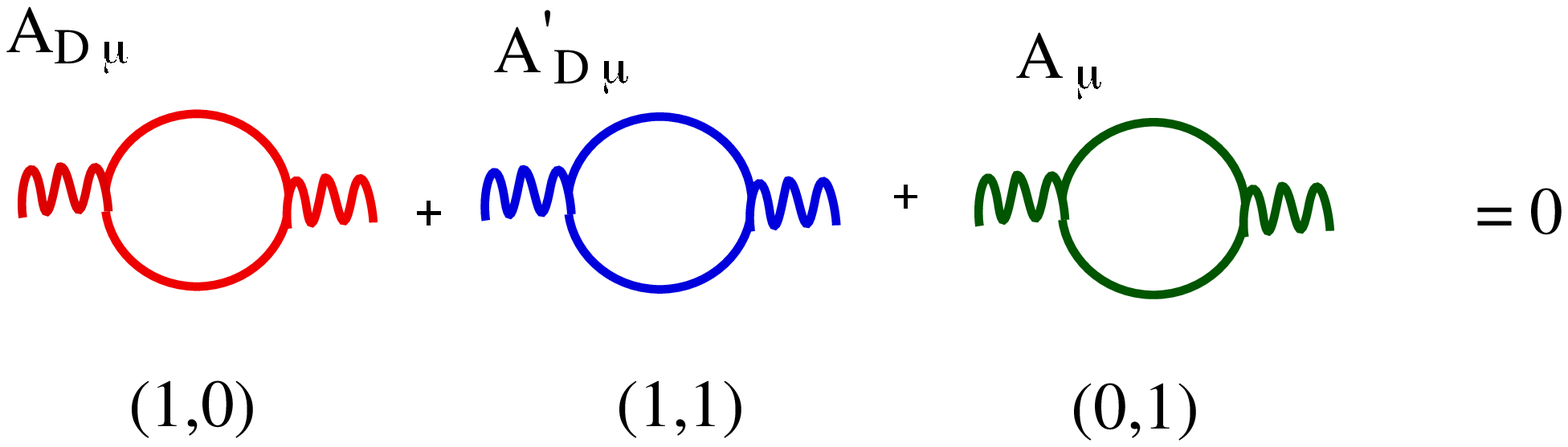}                      
\end{center}
\label{susy}
\end{figure}

Now how do  these massless particles give a vanishing beta function?    In the case of a 
nontrivial  $U(1)$  IR fixed point of  the pure ${ N}=2,$   $SU(3)$  Yang-Mills  theory,
cancellation occurs among a  monopole, a dyon and an electron\cite{AD} (see Figure above).

   The cancellation of $b_0$  in our  case (consider $U_1(1)\subset SU(2)$) is more involved since now there  are  also contributions  of the
gauge multiplet.   Nonetheless, 

\begin{enumerate}

\item four monopole doublets $(\mp 1,0)^4$    cancel the contribution  of the gauge multiplets;
\item a dyon doublet $(\pm 2, \pm1)$  and an electric doublet   $(0, \pm1)$ cancel  each other as  
$$\sum_i (q_i+m_i
\tau)^2  =  1 + (2\, \tau  +1)^2=0, \qquad   {\hbox{\rm  for }}    \quad     \tau^*=\frac{-1+i}{2}:  $$
showing a nice (non-Abelian) generalization of Argyres-Douglas'   mechanism;  
\item   in the second section cancellation occurs because both the charges and the coupling constant $\tau^{*}$
get transformed   by  $ p_1=\pmatrix{-1& 4 \cr 0 & -1 }$,   and the above argument works for  
  $(\pm 4, \mp1)$ 
and   $(\pm 2,  \mp 1)$   with  
$\tau^*=    { 3 + i
\over10}$!   
This strengthens  our idea that different sections  are simply  different descriptions of the same physics. 

\end{enumerate}

   {\it  Thus the  low-energy theory is an interacting SCFT  with
$SU(2) \times U(1)$  gauge group and 
four  magnetic monopole doublets, one dyon doublet and one electric doublet. }

\subsection   {    Six Colliding ${ N}=1$  Local   Vacua:   }

Another way to study  our   SCFT  would be to consider first the theory with unequal quark masses
and then to take the limit of the equal mass.   The SCFT  singularity splits to six  singularities.  

\begin{itemize}  
\item  Each of the six  ${ N}=1$ theories    is  a local $U(1)^2$  theory with  a pair of massless  Abelian monopoles  $M_i, {\tilde
M}_i,$    ($i=1,2$)  carrying each unit charge with respect to one of the $U(1)$ factors; 
altogether there are  12  massless hypermultiplets (as in the SCFT);   
\item   The effect of ${ N}=1$ perturbation   $\mu \, {\hbox{\rm Tr}}  \, \Phi^2 $  can be studied in a  well-known way,   in
terms of  an effective superpotential:    
\begin{equation}  { P} =   \sum_{i=1}^2   \sqrt {2}   \, A_{D_i}  \,   M_i  {\tilde M}_i  +  \mu \,  U (A_{D1}, A_{D2})  + {\hbox {\rm mass terms }} 
\end{equation}
which leads to      $\langle M_i\rangle\ne 0$,   $\langle  {\tilde M}_i\rangle\ne 0$  (Confinement);    
\item  However,      in the     $m_i \to m $   (SCFT)   limit,  the VEVS of the Abelian monopoles   are  found to vanish: 
$$\langle M_i\rangle \to 0, \quad \langle  {\tilde M}_i\rangle \to 0.   $$  
Analogous phenomenon was found in $SU(2)$, $n_f=1$   theory \cite{GVY}.  

\item  We do know (from  the large $\mu$ analysis, vacuum counting, and holomorphic dependence of physic on $\mu$) \cite{CKMP} however 
that   the flavor group {\it is}   dynamically broken in the perturbed  SCFT vacua:
\begin{equation}   G_F =  SU(4) \times U(1)  \Rightarrow    U(2) \times U(2);    \end{equation}     
 then what is the order parameter   of the symmetry breaking?  

\end{itemize}    

{\it   We propose that   condensation of  $SU(2)$   doublets  
 ${
M}_{\alpha}^i,$   ($\alpha=1,2$,   $\, i=1,\ldots, 4$)  
\begin{equation}    \langle    { M}_{\alpha}^i  \, { M}_{\beta }^j \rangle  =     \epsilon_{\alpha   \beta}  \,   C^{ij} \ne 0, 
 \end{equation}    
is formed  due to       the  strong   $SU(2)$  interactions.}      This is compatible with the known dynamical symmetry breaking
pattern.   Note that,  in the sense of complementarity,   such VEVS can alternatively  be understood as 
$$  \langle    { M}_{\alpha}^i  \rangle  = \delta_{\alpha}^i \, v \ne 0 
$$
i.e.,  color-flavor diagonal  VEVS as in the generic $r$-vacua.

\subsection{Summary }

\bigskip

\noindent{ Softly broken ${ N}=2,$   $SU(n_c)$ gauge theories with  $n_f$  quarks thus  exhibit   
various  confining vacua with:  }

\begin{itemize} 

\item   physics quite different  for 

(i)  $r =0,1$ $\Rightarrow$   Weakly coupled Abelian monopoles;

(ii)  $r<{n_f \over 2}$    $\Rightarrow$   Weakly coupled non-Abelian monopoles;

 (iii)   $ r={n_f  \over2}$    $\Rightarrow$   Strongly  coupled non-Abelian monopoles;

\item   nonetheless, both  at  generic  $r$ - vacua  and  at the SCFT ($r={n_f \over 2}$) vacua,  the non-Abelian monopoles
condense as  
$$ \langle    { M}_{\alpha}^i  \rangle  = \delta_{\alpha}^i \, v \ne 0, \qquad  (\alpha=1,2, \ldots, r; \quad    i=1,2, \ldots, n_f)     $$
(``Color-Flavor-Locking");  

\item  Abelian  and non-Abelian monopoles  apear to be related  as  
 $$   \epsilon^{{\alpha_1} {\alpha_2} \ldots
{\alpha_r} } { M}_{\alpha_1}^{i_1}    { M}_{\alpha_2}^{i_2}   \ldots   { M}_{\alpha_r}^{i_r}
\sim  ``U(1)"  {\hbox{\rm monopole}}. $$  

\end{itemize}

\section{QCD} 

Finally let us come back briefly to the real-world  QCD.  Here

\begin{enumerate}  

\item   no dynamical Abelianization is known to occur; 
    
\item    on the other hand,   in QCD  with $n_f$  flavor,     the original and dual beta functions have the first coefficients
($n_c=3$, ${\tilde n}_c=2,3$)
$$  b_0=   - 11 \, n_c +  2 \,n_f \quad  {\hbox {\rm vs}} \quad   {\tilde b}_0  =  - 11 \, {\tilde n}_c +   n_f:    $$
they have the same sign  because of the large coefficient  in front of the color  multiplicity  ({\it cfr.} Eq.(\ref{betafund})).  

\end{enumerate}

Barring that higher loops change the situation, this leaves us with  the option of   strongly-interacting non-Abelian monopoles.  
Is it possible that non-Abelian monopoles (perhaps certain  composite theirof)  carrying nontrivial flavor $SU_L(n_f)\times SU_R(n_f)$  quantum
numbers condense 
yielding  the global   symmetry breaking such as 
$$   G_F = SU_L(n_f)\times SU_R(n_f) \Rightarrow   SU_V(3),   $$
observed in Nature? 
Are  't Hooft's Abelian monopoles    in some sense  composites of these non-Abelian  monopoles ?

\section*{Acknowledgments}
The author thanks R. Auzzi, S. Bolognesi, G. Carlino, R. Grena, P.S. Kumar,
 H. Murayama and L. Spanu for fruitful collaboration.
The author  also thanks the organizers of the 2002 International Workshop
``Strong Coupling Gauge Theories and Effective Field Theories (SCGT 02)" (Nagoya, November 2002)
where this talk was given, and of  ``Institue of Physics Meeting" (London, February 2003) where a
 slightly modified version of the 
talk was presented, for  stimulating  discussions.


\begin{thebibliography}{100}


\bibitem{TH}  G. 't Hooft, {\bf  Nucl. Phys.   B190}   (1981) 455;
S. Mandelstam,  {\bf Phys. Lett.  53B }  (1975) 476.  

\bibitem{SW1}
N. Seiberg and E. Witten,   {\bf   Nucl. Phys. B426} (1994) 19; Erratum
\textit{ibid.}     {\bf   Nucl.Phys.   B430} (1994) 485, hep-th/9407087.

\bibitem{SW2}
N. Seiberg and E. Witten, {\bf Nucl. Phys.  B431} (1994) 484,
   hep-th/9408099.


\bibitem{curves}

P.~C.~Argyres and A.~F.~Faraggi, {\bf Phys. Rev. Lett {\bf 74}} (1995)
3931, hep-th/9411047;
A. Klemm, W. Lerche, S. Theisen and S. Yankielowicz,  {\bf  Phys. Lett.
{\bf B344} }    (1995) 169, hep-th/9411048;
 {\bf  Int. J. Mod. Phys. A11}   (1996) 1929, hep-th/9505150;
A. Hanany
and Y. Oz,   {\bf  Nucl. Phys. {\bf B452} }   (1995) 283,
hep-th/9505075;
P.  C.  Argyres, M.  R.  Plesser and A.  D.  Shapere,   {\bf  Phys.  Rev.
Lett.  {\bf 75} }      (1995) 1699, hep-th/9505100;
P. C. Argyres and A. D. Shapere,   {\bf    Nucl. Phys. {\bf B461} }    (1996)
437,
hep-th/9509175;   A. Hanany,
 {\bf   Nucl.Phys. {\bf B466}}    (1996) 85,  hep-th/9509176.

\bibitem{ArPlSei}
P. C. Argyres, M. R. Plesser and N. Seiberg, Nucl. Phys. {\bf B471}
(1996)
159, hep-th/9603042.


\bibitem{CKMP}  G. Carlino, K. Konishi and H. Murayama,
 {\bf    Nucl. Phys.  B590}  (2000) 37,     hep-th/0005076; 
G. Carlino, K. Konishi, P. S.  Kumar  and H. Murayama,
 hep-th/0104064,   {\bf    Nucl. Phys.  B608}  (2001) 51.



\bibitem{BK}   S. Bolognesi and K. Konishi,  {\bf    Nucl. Phys.  B645 }  (2002) 337, hep-th/0207161. 


\bibitem{CDyons} A. Abouelsaood, {\bf Nucl. Phys. B226} (1983) 309; P. Nelson and A. Manohar,  {\bf Phys. Rev. Lett. 50}
(1983) 943; A. Balachandran et. al.,   {\bf Phys. Rev. Lett. 50}
(1983) 1553;  P. Nelson and S. Coleman,  {\bf Nucl. Phys. B227} (1984) 1;  E. Weinberg,  {\bf Phys.Rev.D54} (1996) 6351,
hep-th/9605229. 



\bibitem{AGK}  R. Auzzi, R. Grena and K. Konishi,  {\bf    Nucl. Phys.  B653 } (2003) 204, hep-th/0211282.


\bibitem{GNO}   P. Goddard, J. Nuyts and D. Olive,   {\bf Nucl. Phys.  B125}
(1977) 1,    E. Weinberg, {\bf Nucl. Phys. B167} (1980) 500;  {\bf Nucl. Phys. B203} (1982) 445. 

\bibitem{AD}
P. C. Argyres and M. R. Douglas,   {\bf    Nucl. Phys.  B448}  (1995) 93,   hep-th/9505062.


\bibitem{SCF}
P. C. Argyres,  M. R. Plesser,  N. Seiberg and E. Witten, {\bf  Nucl. Phys.
{\bf 461}}   (1996) 71, hep-th/9511154.

\bibitem{Eguchi}
 T. Eguchi,  K. Hori, K. Ito and S.-K. Yang, {\bf  Nucl. Phys. {\bf B471}}
(1996)
430, hep-th/9603002.

\bibitem{lebowitz}
A. Lebowitz, {\bf Israel J. Math. 12} (1972) 223.

\bibitem{GVY} A. Gorsky, A. Vainshtein and  A. Yung,   {\bf  Nucl. Phys.   B584}   (2000) 197,  hep-th/0004087. 

\end{thebibliography}
\end{document}